\documentclass[twocolumn,preprintnumbers,nofootinbib]{revtex4-1}
\usepackage{graphicx}
\usepackage{amssymb}
\usepackage{color}
\usepackage{enumerate}
\usepackage{amsmath}
\usepackage{slashed}
\usepackage{comment}
\def\beq{\begin{equation}}
\def\eeq{\end{equation}}
\def\beqn{\begin{eqnarray}}
\def\eeqn{\end{eqnarray}}

\renewcommand{\texttt}{{}}
\newcommand{\be}{\begin{eqnarray}}
\newcommand{\ee}{\end{eqnarray}}

\begin{document}

\title{
{Finite Entanglement Entropy of Black Holes
}}

\author{Stefano Giaccari} 
\affiliation{Theoretical Physics Division of Particles and Fields, 
                    Faculty of Science, University of Zagreb
                    Bijeni\v{c}ka 32, HR-10000 Zagreb, Croatia}

\author{Leonardo Modesto}
\affiliation{Department of Physics, Southern University of Science and Technology, Shenzhen 518055, China}

\author{Les\l aw Rachwa\l{}}

\author{Yiwei Zhu}

\affiliation{Center for Field Theory and Particle Physics and Department of Physics, 
Fudan University, 200433 Shanghai, China}

\begin{abstract}\noindent
We compute the area term contribution to black holes' entanglement entropy (using the 
conical technique) for a class of local or weakly non-local super-renormalizable gravitational theories coupled to matter. For the first time, we explicitly prove that all the beta functions in the proposed theory, except for the cosmological constant, are identically zero in cut-off regularization scheme and not only in dimensional regularization scheme. In particular, we show that there is no divergence quadratic in cut-off and hence there is no contribution to the beta function of the Newton constant.  As a consequence of this result, we argue that in these theories of gravity conical entropy is a sensible definition of physical entropy, in particular, it is positive-definite and gauge-independent. On top of this the conical entropy, being expressed only in terms of the classical Newton constant, turns out to be finite and naturally coincides with Bekenstein-Hawking entropy. Finally, we propose a theory in which the renormalization of the Newton constant is entirely due to the Standard Model matter, arguing that such a contribution does not give the usual interpretational problems of conical entropy discussed in the literature. 
\end{abstract}

\maketitle






\section{Introduction}

People have long been involved in understanding a big issue of Einsteinian gravity, which is actually  
common to all generally relativistic theories of gravity, namely: what is the nature of Bekenstein-Hawking black hole entropy? There are two possible interpretations of the famous entropy formula $S=A/4$.
It can have a statistical mechanics origin or, given the black hole state, a quantum entanglement interpretation.
If we believe in a statistical origin we should be able to identify the microscopic degrees of freedom 
compatible with the macroscopic area law. This was achieved in string theory by Strominger and Vafa \cite{Strominger:1996sh} with a very educational explicit computation. However, 
it is not entirely clear what the role of the black hole event horizon is in this context. 
Indeed, what is generically relevant in this approach is the correct identification of the gravitational 
source, which is assumed to be located at the singularity point because usually the Ricci tensor is zero everywhere else. 
Therefore, for some reason the matter inside a black hole must undergo a peculiar statistical mutation 
during the gravitational collapse that is not displayed in any other condition. 

On the other hand, in the entanglement interpretation of the black hole entropy 
the event horizon is just a particular boundary surface splitting the Hilbert space in a 
tensor product of two Hilbert spaces for the external and internal regions.  However, the physical interpretation in this case seems to be quite elusive because the entanglement entropy evaluated by the so called ``replica trick"  for a generic quantum field theory is typically divergent. It was proposed by Susskind and Uglum \cite{Susskind:1994sm} that the UV divergences in the area term of the entanglement entropy could be absorbed in a renormalization of the gravitational coupling. This proposal has been discussed in a large number of papers (see \cite{SoloLiving} for a review), being confirmed in some cases, but not in others.


%

 In this paper we mainly deal with  the so-called {\em conical entropy}, which has been discussed in the literature \cite{SoloConi, SoloStelle} as a way to get an entropy from the replica trick and whose renormalization coincides with the one expected for the Wald entropy of black holes as far as the area term is concerned.  Similarly to entanglement entropy, it is evaluated applying the Callan-Wilczek formula \cite{Larsen:1995ax, Callan:1994py} to the gravitational quantum effective action $W$ on a general regular background. Afterwards, such background is deformed to get the effective action $W(\alpha)$ for the $\alpha$-fold covering {$E_\alpha$}. There exists a standard procedure to relate the curvature terms computed on a smooth manifold {$E$} to the corresponding ones for a conifold {$E_\alpha$} \cite{SoloLiving, SoloStelle, SoloRecent, SoloConi}. 
 Even if the conical entropy has the attractive feature of reproducing the expected area term of Bekenstein-Hawking entropy in terms of the renormalized gravitational constant, in general it cannot be given a consistent statistical interpretation. In fact, the surface term of the effective action, which gives the area term of the conical entropy, will in general receive UV-divergent contributions which are gauge dependent and negative (in particular from gauge vector bosons and gravitons \cite{SoloRecent}). However,  in the context of super-renormalizable gravitational theories \cite{Krasnikov, Tombo, Khoury, modesto,modestoLeslaw, universality, Briscese:2013lna, Cnl1, Dona,Mtheory,Modesto:2013jea}, the beta functions are known to be gauge-independent  and furthermore they are completely determined at one loop \cite{shapiro1}. On top of this, it has been explicitly shown in ref. \cite{modestoLeslaw} that these beta functions can be fixed to zero by including in the action a finite number of operators (leading on a flat background only to vertices) whose couplings are completely determined by a one-loop computation. Therefore, the  renormalized gravitational constant $G_N$ can be chosen, by a completely gauge invariant renormalization procedure, as a positive quantity. In result, the Bekenstein-Hawking entropy in such theories gets a statistical interpretation from its identification with the conical entropy. Moreover, manifestly generally covariant, always positive (since the leading area term is proportional to the renormalized Planck mass scale), and UV-finite entropy is to be interpreted as a viable candidate for a quantum entropy of black holes. 
 Contrary to entanglement entropy, conical entropy takes into account the back-reaction of operators with non-minimal couplings. The latter contain curvature and due to the conical singularity of an $\alpha$-fold covering an imprint of curvature of spacetime is to be seen in the computation of the entropy. This property is actually crucial in making the conical entropy a reliable probe of the UV-finiteness of higher derivative or non-local theories. 


In \cite{Cooperman:2013iqr} some limitations to this interpretation were pointed out. 
One is related to the fact that, if the entangling surface has non-vanishing extrinsic curvature in the time slice, the standard  renormalization procedure for entropy fails. In this paper we will only consider entangling surfaces fulfilling the required condition for this renormalization procedure to be applied. 
Moreover, we will assume that a fully generally covariant regularization of quantum gravity (involving the quantum fluctuations of the metric itself) can be carried out even in the presence of conical singularities. On the other hand both the dependence on the regularization scheme and on an arbitrarily chosen renormalized Planck scale are problems that should be automatically fixed in the framework of a consistent UV completion of quantum gravity. In string theory, in particular, several authors (see, for example,  \cite{Dabholkar:1994gg, Dabholkar:1994ai, Dabholkar:2001if}) have argued that the entanglement entropy should turn out to be finite as a consequence of the natural UV cut-off provided by the string length. Recent computations in these directions have been done either using a slightly modified definition of conical entropy \cite{He:2014gva} or, in the context of two-dimensional string theory, dual to some matrix quantum models, in the semi-classical limit of weak string coupling \cite{Hartnoll:2015fca}.  A fully conclusive computation in a generic setup is still missing. Of course, it would also be of the utmost interest to check the possibility of getting a finite entropy in a purely quantum field 
theoretical framework \cite{Paddy}. If entropy is directly related to the counting of the classical degrees of freedom in the field theory (or microstates in quantum theory), then of course infinities are to be expected. This seems to be the obvious expectation in asymptotically-free theories where interactions die off in the UV regime. On the other hand, for super-renormalizable or UV-finite theories where interactions are crucial in determining the UV behaviour, we may expect that a proper definition of entropy should take these interactions into account.
 Actually, even if the power structure of divergences contributing to entanglement entropy changes with the interactions \cite{Hung:2011xb}, few examples of straightforward computations for interacting theories on some conical manifolds are known (see \cite{Casini:2014yca} for instance). In this paper we want to bring about such a task of deriving the conical entropy in the case of a proposed class of UV-complete theories of quantum gravity.
 
Another puzzle is the one related to the various definitions of black hole entropy.  In the field-theoretical framework the statistical entropy counts the number of degrees of freedom, therefore, it is naturally divergent if we deal with continuous fields. 
The same happens for the entanglement entropy of a black hole horizon. However, this is in strong disagreement with the finite, non-divergent results, for the entropy of black holes computed using the classical Wald formula applied to the quantum effective action of gravitational fluctuations. Furthermore, entanglement entropy is in general positive-definite, whereas the Wald entropy is not, as it lacks a general statistical interpretation. On the other hand, entanglement entropy seems to be quite insensitive to the non-minimal couplings in the classical gravitational theory which we start from. Instead, these couplings are crucial in the computation of the quantum effective action used in the definition of Wald entropy and are also essential in making the theory UV-finite.  Hence it is necessary to understand how to make the entanglement entropy finite in the quantum field theory framework. Our main motivation for this work was actually the idea that in the context of a super-renormalizable or  finite quantum field theories the relation between these two seemingly very different objects can actually turn out to be clearer. The results presented in this paper actually show that, in the case of super-renormalizable or finite theories, conical entropy becomes a fully physical object coincinding with Wald entropy.

The present paper is therefore organized as follows.
In section II we briefly introduce
a class of weakly non-local 
theories of gravity, which are unitary (ghost-free) and perturbatively super-renormalizable or finite in the quantum field theory framework
\cite{Krasnikov, Tombo, Khoury, modesto,modestoLeslaw, universality, Briscese:2013lna, Cnl1, Dona,Mtheory,Modesto:2013jea}. 
At classical level evidences endorse that we are dealing with 
``{\em singularity-free gravitational}\hspace{0.05cm}" theories 
\cite{ModestoMoffatNico,Frolov:2015bta,BambiMalaModesto2, BambiMalaModesto,calcagnimodesto, koshe1} (see also the recent papers \cite{Buoninfante:2018xiw, Koshelev:2018hpt}). 
 However, the Einstein spaces seem still to be exact solutions of the non-local theory
\cite{exactsol, BrisceseClassical}, although it is still a debated open problem what kind of energy tensor could source such spacetimes in a non-local theory \cite{appear}. 
Nevertheless, the whole analysis in this paper only needs the presence of an event horizon regardless of the spacetime structure at short distance. Therefore, the analysis can be applied to singular as well as singularity-free black holes. 
In section III we discuss in detail how to achieve super-renormalizability and finiteness of the theory in both dimensional (DR) and cut-off regularization schemes. 
In section IV we take the Callan-Wilczek formula \cite{Callan:1994py} as the operational definition of renormalized conical
entropy. 
In section V we  compare the renormalized conical
entropy with the Wald entropy formula for black holes \cite{Wald:1993nt,Iyer:1994ys} in the gravitational theories under consideration, finding that the area law terms coincide in the two cases. 
%
%
We conclude by summarizing our results that 
can be generalized to any local higher derivative or non-local gravitational theory.
To avoid cumbersome technical details we will often refer to \cite{SoloLiving, SoloRecent}
and references within. 


\section{General theory}
The most general $D$-dimensional theory weakly non-local (or quasi-local) and quadratic in the Riemann curvature reads 
\cite{Krasnikov, Tombo, Khoury, modesto,modestoLeslaw,universality, Briscese:2013lna, Cnl1, Dona, Mtheory, 
Modesto:2013jea, M3,M4,kuzmin}, 
\be
\label{theory}
&& \hspace{-0.3cm}
\mathcal{L}_{\rm g} = -  2 \kappa_{D}^{-2} \, \sqrt{|g|}\left[ \,   R \, +  \label{gravityG}  R \, 
 \gamma_0(\Box)
  R \right. \\
&& \hspace{-0.3cm}
\left.  
 + {\bf Ric} \, 
\gamma_2(\Box)
 {\bf Ric}
+ {\bf Riem}  \, 
\gamma_4(\Box)
{\bf Riem} 
+  V_K \, 
\right]  .
\nonumber 
\ee
%
%
%
%
The above expression of the Lagrangian of the theory will be particularly suitable for the evaluation of Wald entropy.
The action of the theory consists of a kinetic weakly non-local operator quadratic in the curvature, three entire functions 
$\gamma_0(\Box)$, $\gamma_2(\Box)$, $\gamma_4(\Box)$, and typically local terms in $ V_K$ which are at least cubic in the curvature tensor, namely $ V_K \sim O(\mathcal{R}^3)$. 
 %
Some of the operators in ${V_K}$ are called killers because they are crucial in making the theory finite in any dimension. 
Moreover, 
$\Box = g^{\mu\nu} \nabla_{\mu} \nabla_{\nu}$ is the covariant box operator. Next, an integer $\rm{N}$ is defined to be the following function of the spacetime dimension $D$: $2 \mathrm{N} + 4 = D$. The coupling constant $\kappa_D^{-2}$ is related to the Newton constant via $\kappa_{D}^{-2}=1/(32 \pi G_N)$.
The form-factors $\gamma_i(\Box)$ must take the following particular form if we require the same spectrum as in Einsteinian gravity. We write them in terms of exponentials of entire functions $H_\ell(z)$ ($\ell=0,2$), namely 
\be
&&
\hspace{-0.5cm} \gamma_0(\Box) = - \frac{(D-2) ( e^{H_0} -1 ) + D ( e^{H_2} -1 )}{4 (D-1) \Box} + \gamma_4(\Box) 
\label{gamma0} \, , \\ 
&&  \hspace{-0.5cm}
 \gamma_2(\Box) = \frac{e^{H_2} -1 }{\Box} - 4 \gamma_4(\Box) \, .
\label{gamma2}
\ee
The form-factor $\gamma_4(\Box)$ stays arbitrary, but is constrained by renormalizability to have the same (or lower-power) asymptotic UV behaviour as the other two form-factors $\gamma_\ell(\Box)$ ($\ell=0,2$). The minimal choice 
compatible with unitarity and super-renormalizability corresponds to  $\gamma_4(\Box) =0$.

%

Finally, the entire functions $V^{-1}_{\ell}(z) \equiv \exp H_{\ell}(z)$ ($z \equiv - \Box_{\Lambda} \equiv - \Box/\Lambda^2$) ($\ell=0,2$) introduced in \eqref{gamma0} and \eqref{gamma2}
are real and positive on the real axis and without zeros on the 
whole complex plane $|z| < + \infty$. (Here $\Lambda$ is an invariant mass scale in our fundamental theory defining the so-called scale of non-locality.) The last requirement implies that there are no other
gauge-invariant poles than the transverse massless physical graviton pole. Moreover,  
%
there exists an angle $\Theta$ ($0<\Theta<\pi/2$), such that asymptotically
\be
&& \hspace{-1cm} 
|V^{-1}_{\ell}(z)| \rightarrow | z |^{\gamma + \mathrm{N}+1} \quad {\rm for }\quad  |z|\rightarrow + \infty
, \quad 
\gamma\geqslant \frac{D}{2} \, , 
\label{tombocond}
\ee
for the complex values of $z$ in the conical regions $C$ defined by: 
$C = \{ z \, | \,\, - \Theta < {\rm arg} z < + \Theta \, ,  
\,\,  \pi - \Theta < {\rm arg} z < \pi + \Theta\}.$
The last condition is necessary to achieve the maximum convergence of the theory in
the UV regime avoiding non-local counterterms.  One example of such function, due to Tomboulis \cite{Tombo}, is:
\be
V^{-1}(z)= e^{1/2 \left[ \Gamma \left(0, p(z)^2 \right)+\gamma_E  + \log \left( p(z)^2 \right) \right] } ,
\label{TomboFF}
\ee
where $p(z)$ is a polynomial of degree $\gamma+{\rm N} +1$. Above $\Gamma(a,z)$ stands for the incomplete Gamma function and $\gamma_E$ is the Euler-Mascheroni constant. In most of the analysis below we will assume that this UV polynomial $p(z)$ is actually a monomial $z^{\gamma+{\rm N} +1}$.

\paragraph*{Propagator and unitarity ---}
Splitting the spacetime metric into the flat Minkowski background and the perturbation $h_{\mu \nu}$ 
defined by $g_{\mu \nu} =  \eta_{\mu \nu} + \kappa_D \, h_{\mu \nu}$,
we can expand the action (\ref{gravityG}) to the second order in $h_{\mu \nu}$.
The result of this expansion together with the usual harmonic gauge fixing term reads \cite{HigherDG}
$\mathcal{L}_{\rm quad} + \mathcal{L}_{\rm GF} = h^{\mu \nu} \mathcal{O}_{\mu \nu, \rho \sigma} \, h^{\rho \sigma}/2$,
where the operator 
$\mathcal{O}$ is made up of two terms, one coming from the quadratization of (\ref{gravityG})
and 
a gauge-fixing term
\cite{Stelle, Shapirobook}.
The d'Alembertian operator in $\mathcal{L}_{\rm quad}$ and the gauge fixing term are written in terms of  
flat spacetime metric and derivatives.
Inverting the operator $\mathcal{O}$ \cite{HigherDG} and making use of the
form-factors (\ref{gamma0}) and (\ref{gamma2}), we find the 
two-point function in the harmonic gauge ($\partial^{\mu} h_{\mu \nu} = 0$),
\be
 \mathcal{O}^{-1} = 
\frac{P^{(2)}}{k^2   e^{H_2(k^2/\Lambda^2)} }
\label{propagator} 
- \frac{P^{(0)}}{  \left( D-2 \right)k^2   e^{H_0(k^2/\Lambda^2)}} \,  .
\ee
%
 Here we omitted gauge-dependent terms and the tensorial 
indices on the propagator $\mathcal{O}^{-1}$. The usual projectors $\{ P^{(0)},P^{(2)}\}$ 
are defined in  
\cite{HigherDG, VN}\label{proje2}.
%
We also have replaced $-\Box \rightarrow k^2$. 

The propagator (\ref{propagator}) describes the most general spectrum compatible with unitarity without any other degree of freedom besides the massless spin-$2$ graviton field. Indeed the optical theorem is trivially satisfied, namely 
 \be
2{\rm Im} \left\{  T(k)^{\mu\nu} \mathcal{O}^{-1}_{\mu\nu, \rho \sigma} T(k)^{\rho \sigma} \right\} >0, 
\ee
 where $T^{\mu\nu}(k)$ is the most general conserved energy-momentum tensor in momentum space. 
%
%
%

The tensorial structure in (\ref{propagator}) is the same as in Einsteinian gravity, but the multiplicative 
 factors $\exp H_\ell(-\Box_{\Lambda})$ for $\ell=0,2$ make the theory strongly UV-convergent without the need to modify the spectrum or introducing ghost instabilities. 
The detailed reference about unitarity, super-renormalizability, and UV-finiteness issues in non-local theories around the Minkowski spacetime can be found in \cite{Krasnikov, Tombo, Khoury, modesto,modestoLeslaw,universality, Briscese:2013lna, Cnl1, Dona, Mtheory,
Modesto:2013jea, M3,M4}. 
 Moreover, recently it has been proved that a slight modification of the theory is stable around any maximally symmetric spacetime \cite{StableDeSitter1,StableDeSitter2,KKMR}.

\section{Super-renormalizability and finiteness}

\subsection{
Power counting} 
 %
We now review the power counting analysis of the quantum divergences. Additionally, in the next section we will make an important distinction between truly polynomial and monomial UV behaviour of the theory because in the last case we have less divergences. But for the moment we remain very general.

In the high energy regime, 
the above propagator (\ref{propagator}) in momentum space 
scales schematically  as: 
$\mathcal{O}^{-1}(k) \sim k^{- (2 \gamma +D) }$.  
The vertices can be collected in different sets that may or may not involve the entire functions $\exp H_\ell(z)$. 
However, to find a bound on the quantum divergences it is sufficient to concentrate on
the leading operators in the UV regime. 
These operators scale as the propagator giving the following 
upper bounds on the superficial degree of divergence of any graph $\omega(G)$ \cite{modesto, shapiro3, HigherDG0, betaGN}, 
\be
&& \delta^D(K) \, \Lambda^{2 \gamma(L-1)} \int (d^D p)^L \left( \frac{1}{p^{2 \gamma+D}} \right)^I \left( p^{2 \gamma +D} \right)^V \nonumber \\&&
 =
\delta^D(K) \, \Lambda^{2 \gamma(L-1)} \int (d^D p)^L \left( \frac{1}{p^{2 \gamma+D}} \right)^{L-1} \nonumber \\
&& 
= \delta^D(K) \, \Lambda^{2 \gamma(L-1)} \, \left( \Lambda_{\rm cut-off}\right)^{\omega(G)} \,  , \nonumber \\
&& 
\boxed{\omega(G) \equiv D - 2 \gamma  (L - 1) } \,\, ,
\label{even}
\ee
%
where we introduced the following notation: $V$ for the 
numbers of vertices, $I$ for internal lines, $L$ for the 
number of loops, $K$ for the sum of external momenta, $\Lambda_{\rm cut-off}$ for the cut-off scale. We also used the topological relation: $I = V + L -1$. 
Thus, if $\gamma > D/2$, only 1-loop divergences survive.  
Therefore, 
the theory is super-renormalizable \cite{Krasnikov, Efimov}
and only a finite number of operators of mass dimension up to $D$ has to be
included in the action in even dimension.  For the sake of simplicity, we presented this result assuming a flat Minkowski background metric, but it can be generalized to a generic background (in particular to one involving an event horizon) using the standard background field method. On the other hand, recently a similar result has been proven for gravity on the (A)dS background \cite{KKMR}.  We also remind that UV-divergences are independent on the background because in the UV limit every smooth manifold is flat. Physically speaking, these divergences probe the spacetime structure when two points get to coincide with each other.

\subsection{Divergences in dimensional regularization scheme}

Let us first consider the divergences of the theory in dimensional regularization \cite{GBV}. 
In this scheme if the asymptotic behaviour of the form-factors $\exp H_\ell$ is monomial 
and the integer $\gamma$ satisfies the constraints of the previous section, 
then only the beta functions for the operators $\cal O$ of dimension $D$ are non-zero, namely 
\be
\hspace{-0.4cm}
\beta_{{\mathcal O}^{\left[D\right]}} \neq 0  , \,\, \beta_{{\mathcal O}^{\left[D-2\right]}} = 0 , \,\,
 \beta_{{\mathcal O}^{\left[D-4\right]}} = 0 , \,\, \dots , \,\,  
 \beta_{{\mathcal O}^{\left[0\right]}} = 0 .
 \label{cumbersome}
\ee
%
This is due to the fact that in DR scheme we do not have any additional mass scale parameter and the coefficients of covariant divergent terms must be dimensionless (for form factors asymptotically monomial).
For the sake of simplicity we here consider the minimal four-dimensional theory compatible with
unitarity, which is moreover sufficient to obtain finiteness in dimensional regularization, namely:
\be
&& \hspace{-0.8cm}
\mathcal{L}_{\rm g} = - 2 \kappa_4^{-2}  \Big( R 
 + G_{\mu \nu} \,  \frac{ e^{H(-\Box_{\Lambda})} -1}{  \Box}   R^{\mu \nu}  \nonumber \\
 && \hspace{0.1cm}
+   s_1 \, R^2 \, \Box^{ \gamma -2} R^2 + s_2 \, 
R_{\mu \nu} R^{\mu\nu} \, \Box^{ \gamma -2} R_{\rho \sigma} R^{\rho \sigma}  \Big) .
\label{eq:NLfourdim}
\ee
Generalizations to extra dimensions are straightforward \cite{modestoLeslaw}. 
Let us assume $s_1=s_2 = 0$ for the moment. 
For asymptotically {\em monomial} form-factors, one-loop divergent contributions 
can come from vertices generated only by the form-factors 
while the Einstein-Hilbert $\sqrt{|g|}R$
term does not produce any divergence (in both dimensional and cut-off regularization scheme). Indeed, the propagator in the ultraviolet regime falls off much faster than the scaling behaviour of the vertices coming from  the two-derivative
term above. 
The counterterms are proportional to $R^2$ and $R_{\mu\nu}^2$ only, so
the only non-zero beta functions in $D=4$ are 
$\beta_{{ R}^2 } \neq 0$ and $\beta_{\rm Ric^2} \neq 0$, whereas 
$\beta_R\equiv\beta_{ G_N} =0$.
It is to be noticed that, as $ \beta_{{\mathcal R}^{0}} = 0$,  no cosmological constant term is produced as a quantum correction. For the aim of this paper it is relevant that there is no renormalization of the Newton constant ($\beta_{G_N}=0$). 
For the case of Minkowski signature, one more reason to use DR is that 
the cut-off regularization scheme is not naively Lorentz invariant (see however \cite{Codello:2015oqa} for a different point of view). 
As we noticed above, 
the cancellation of some beta functions is actually automatically valid for perturbations of gravity around a background that is a classical solution of \eqref{eq:NLfourdim} (in particular a Ricci-flat one). It is also possible to generalize this analysis to the case when a cosmological constant term $\bar\lambda$ is included (see \cite{StableDeSitter1,StableDeSitter2,KKMR}).

\subsection{Divergences in cut-off regularization scheme} 

Let us again focus on \eqref{eq:NLfourdim} to avoid cumbersome operators from \eqref{cumbersome}.
In the cut-off regularization scheme (see the Appendix for an explicit one-loop computation) we expect, besides logarithmic divergences, extra quartic and quadratic ones in a cut-off $\Lambda_{\rm cut-off} \equiv k $. Let us consider in details the case of quadratic divergences, because they appear in the renormalization of the Newton constant. 
Using the heat-kernel expansion the divergent contributions to the quantum effective action are:
\be
&& \hspace{-1cm} 
 \Gamma^{(1)}_{{\rm div} } =  
 \int\! d^4 x \sqrt{|g|} \left[ 
 \left(\beta^{(0)}_{G_N} \log k^2+  \beta^{(2)}_{G_N}  k^2\right) R \right.\nonumber \\
 && \hspace{0.5cm} 
 \left.
+ \left( \beta^{(0)}_{\bar{\lambda}}  \log k^2 + \beta^{(2)}_{\bar{\lambda}}  k^2 + \beta^{(4)}_{\bar{\lambda}} k^4   \right) \right. 
\nonumber \\&&  \hspace{0.5cm} 
\left. 
+ \beta_{R^2} \log k^2 \, R^2 + \beta_{{\rm Ric}^2} \log k^2 \, R_{\mu\nu} R^{\mu\nu}
\right],
\label{eq:DivCutoff}
\ee
where the coefficients in front of each term are related to the beta functions of the corresponding couplings. 
The general structure of quartic, quadratic and logarithmic divergences is displayed. In particular, the beta function for the Newton constant is given by
\be 
\beta_{G_N}=k\frac{d \kappa_4^{-2}}{dk}=\beta^{(0)}_{G_N}  +\beta^{(2)}_{G_N}k^2\, , \,\,\, \kappa_4^2 = 32 \pi G_N  . 
\label{eq:betaGN}
\ee
 Above $\beta^{(0)}_{\bar{\lambda}}$, $\beta^{(2)}_{\bar{\lambda}}$, $\beta^{(4)}_{\bar{\lambda}}$, $\beta^{(0)}_{G_N}$, $\beta^{(2)}_{G_N}$, $\beta_{{R}^2}$ and $\beta_{{\rm Ric}^2}$ are numerical constants depending only on the non-running coupling constants in front of the higher derivative terms. So, in particular, in the case of asymptotically monomial form-factors in UV there are no other sub-leading divergences than the ones already present in \eqref{eq:DivCutoff} with the highest power exponents on the cut-off. Specifically this means that in formula \eqref{eq:betaGN} $\beta^{(0)}_{G_N}=0$. Again the reason is that we cannot form a dimensionful ratio having only  one coupling in front of the leading term in the UV monomial. (In a recent paper \cite{betaGN} the full beta function for the Newton constant in general higher derivative theories has been computed in DR scheme.)
 
Therefore, contrary to what happens in DR, in cut-off regularization scheme we have an infinite renormalization of $G_N$. 
This is very crucial in determining the correct form of the entanglement or conical entropy 
\cite{Nodimreg}. 
However, we can add other operators  to the action, without changing the perturbative spectrum or  
affecting unitarity, in such a way that the beta function for $G_N$ will be vanishing. Useful operators giving contribution to $\beta_{G_N}$ only linear in their front 
coefficients are:
\be \hspace{-0.2cm}
s_a {R}^2 \Box^{\gamma-1} { R} , \,\,  s_b  {\bf Ric}^2 \, \Box^{\gamma-1} { R} , \,\, 
s_c  {\bf Riem}^2 \,  \Box^{\gamma-1} { R} ,  ...  \quad
\label{CoffK}
\ee
When the background field method is employed all the above operators contribute to the beta function of $G_N$ 
linearly in $s_a$, $s_b$, $s_c$, etc, namely 
\be
\beta_{G_N} = k^2\sum_i s_i \times {c}_i \,  + \dots \,  ,
\label{linear_beta}
\ee
where $\dots$ means contributions from other local or non-local operators or the terms in $ V_K$ present in the full theory \eqref{theory}. The coefficients $c_i$ are c-numbers inversely proportional to the coupling constants in front of the highest derivative terms (of the type $\omega_\gamma{\cal R}\square^\gamma {\cal R}$) quadratic in curvature, which result from the UV behaviour of the form-factor. Actually, the numerical coefficients $c_i$ carry energy dimension because of the omega coefficients hidden there.  In total they conspire with the dimensionful parameters $s_i$ making the correct energy square dimension on both sides of the above equation.

To get the above equation we use the background field method, Barvinsky-Vilkovisky trace technology \cite{GBV} for computing traces and the dimensional analysis to constrain the dependence on the parameters $s_i$. It is obvious that the terms in \eqref{CoffK} give contributions only linear in the front coefficients after we take into account dimensional analysis and the expression for the second variation of these operators on a general background. These variations are at least linear in curvature. We are looking for UV divergences according to the formula for the divergent part of the effective action 
\be
\Gamma_{\rm div}=\frac{1}{2}{\rm Tr}_{\rm div}\ln\frac{\delta^2S}{\delta h_{\mu\nu}\delta h_{\rho\sigma}}.
\label{secvar}
\ee
To compute the trace we expand the logarithm in an infinite power series. To get 
$\beta_{G_N}$ we only need to look at the trace of an operator with first power of any curvature and this always comes  linearly in the front coefficients $s_i$ in the second variation.  So in the expression for a divergent part of the trace of the logarithm we find only linear dependence on $s_i$.  Due to the properties of the heat kernel it is clear that the effective action cannot contain other powers of $k$, such as non-even-integer powers, or functions other than logarithms. This proof of linearity is the key point of the paper and a proper attention should be given by the reader to the derivation of the equation \eqref{linear_beta}. In our class of asymptotically polynomial theories the UV divergences are given by the divergences of a local higher derivative (polynomial) theory. Hence, as emphasized in the introduction, here we use methods of heat kernel applied to local higher derivative theories. This is allowed because the UV divergences are the same and they depend only on a UV behaviour of the theory.

Since the equation \eqref{linear_beta} is linear in $s_i$ it can always be solved for one of the coefficients $s_i$. Let us say that this value is $s_{i*}$. If we adjust the coefficient $s_i$ such that $s_i=s_{i*}$, then the beta function for the Newton constant $\beta_{G_N}$ vanishes. Therefore, in this modified theory 
there is no infinite renormalization of the Newton constant. This result is one-loop exact because there are no divergences from two loops upwards. When gravity is coupled to matter we still have 
super-renormalizability if the matter sector is not self-interacting (see \cite{universality}).
However, a 
weakly non-local 
extension of gauge interactions \cite{universality}, together with the fermionic and scalar sector of the standard model of particle 
physics, will be sufficient \cite{Tombo, universality} to achieve super-renormalizability
also for self-interacting matter.

 We remark here that the choice of the adjustable parameter $s_i$ does not influence unitarity at all because one can easily check that the optical theorem is here satisfied for whatever value of the $s_i$ coefficients. Indeed, around the flat spacetime the terms cubic or higher in curvatures do not have any impact on the propagator of the gravitational perturbations. Regarding 
 the positivity of the gravitational energy the sign of $s_i$ may matter on a non-flat background, but this is not an issue of unitarity. Moreover, 
 a non-perturbative definition of unitarity around any non-flat background is a complicated and not fully understood issue.

Notice that in odd dimension there are no one-loop divergences 
in DR, 
because we cannot construct curvature invariant operators with an odd number 
of derivatives of the metric tensor. Moreover, this result is one-loop exact because we do not have divergences for 
$L>1$. 
However, for all the theories here proposed the maximal divergence of the cosmological constant is still present in any dimension when we implement the cut-off regularization scheme.

Finally, the killers needed in cut-off regularization scheme are completely harmless for the beta functions in DR scheme. Indeed, the theory that is UV-finite in cut-off regularization scheme is also automatically finite in DR scheme, but not vice versa. The killers in cut-off scheme are spectators from the point of view of DR. They, however, may give different contributions to the finite pieces of the quantum effective action. For example, in a case of a UV monomial theory we have no divergences proportional to $R$ (which renormalize $G_N$) in DR, but in the cut-off scheme the suitable killer must be added.
Moreover, we emphasize that the UV divergences are independent on the background spacetime. For example, in the next section our analysis will be restricted to UV divergences and finiteness of the entanglement entropy associated to the surface of the black hole horizon.

\subsection{Gauge and matter sectors}
In the papers \cite{modestoLeslaw, universality} weak non-locality has been extended to all fundamental 
interactions. This is an inescapable extension beyond the standard model if we want to preserve 
super-renormalizability of the gravitational interactions after coupling to matter. Moreover, 
the weakly non-local gauge interactions turn out to be (super-)renormalizable or finite regardless of the spacetime dimension. Following the notation of section II, the Lagrangian for gauge bosons reads as follows, 
\be \hspace{-0.2cm} 
{\cal L}_{\rm gauge} = -\frac{1}{4g^2}\Big[ {\bf F} e^{H({\cal D}_\Lambda^2)}
{\bf F} + s_g {\bf F}^2({\cal D}_{\Lambda}^2)^{2} {\bf F}^2 \Big],
\label{gauge}
\ee
where $H$, as a function of the square of the gauge covariant derivative ${\cal D}$, can be chosen to be an entire one having the same asymptotic behaviour as the analogue functions introduced for the pure gravity sector.
For the fermionic and scalar sectors we achieve super-renormalizability with the following Lagrangians,
\be
&&  
\mathcal{L}_{\rm F} = 
 \sum_{a}^{N_f}\bar \psi_a \, i \slashed{\cal D}_{a}  e^{H(\slashed{\cal D}^{2}_{a,\Lambda})}
\, \psi_a \label{FS} \, , \\
&& 
\mathcal{L}_{\rm H} =  ({\cal D}_{\mu} \Phi)^\dagger e^{H({\cal D}_{\Lambda}^2)} ({\cal D}^{\mu} \Phi) 
- \mu^2 \Phi^\dagger  e^{H({\cal D}_{\Lambda}^2)}   \Phi - \lambda (\Phi^\dagger \Phi)^2. 
\nonumber 
\ee
To achieve full finiteness of all running coupling constants we need few other operators. 
However, this goes beyond the scope of this paper. For the interested reader we refer to  \cite{universality, 
brisceseScalar}. We add that in the quantum coupled system where we have gravitation and gauge and/or matter sectors, the beta function for the Newton constant can be made zero
 by the same method as the one used in \eqref{linear_beta}.

\section{
 Conical entropy} 
 \label{s:ConicalEntropy}
%

In this section we consider the conical 
entropy of black hole solutions for the class of theories (\ref{gravityG}) exhibiting perturbative unitarity and ultraviolet finiteness. First we discuss the classical black hole solutions pointing to the fact that for the subsequent analysis only the presence of an event horizon matters. Then we construct a conifold to evaluate the conical entropy there using the Callan-Wilczek method of the effective gravitational action. After this we study the divergences of this conical entropy and explicitly show how to avoid them. The universal statistical interpretation of the UV-finite entropy is also given at the end.

The results in this section can be obtained both in pure gravity and for the case of matter and gauge fields coupled to it. The non-locality of our theory is important only for the unitarity issue (there is a rigorous proof for this in \cite{Tombo, modesto}), while for other aspects related to super-renormalizability, UV-finiteness, conical singularities, Callan-Wilczek formula, applications of the heat kernel methods,  and, for various expansions, the local higher derivative theories are sufficient. These local theories arise as UV limits of non-local theories and we base our analysis of conical entropy on the situation with higher derivative gravitational theories.

Since \eqref{theory} is a modified gravity theory then it can contain black hole solutions 
for which we want to compute the entanglement entropy. 
Notice, that the content of this section is general and independent on the particular solution as long as it shows an event horizon. Therefore, we can for example apply our analysis to any black hole solution, singular \cite{exactsol} or singularity-free 
\cite{ModestoMoffatNico,Frolov:2015bta,BambiMalaModesto2, BambiMalaModesto,calcagnimodesto, koshe1,Buoninfante:2018xiw, Koshelev:2018hpt}. Moreover, as we remarked in the introduction, our results can be easily exported to local higher derivative theories \cite{shapiro3,lm5}, where in the conditions stipulated above, we are sure that the Schwarzschild metric is an exact black hole solution.

We here study uncharged non-rotating black holes described by the Schwarzschild metric.
For such a metric there exists a time-like Killing vector 
$\partial_\tau$, which is null at the horizon surface $\Sigma$. In the vicinity of this bifurcation surface, 
the spacetime is therefore a product of the surface $\Sigma$ and a two-dimensional disk $D_2$. On $D_2$ the time coordinate plays the role of an angular coordinate after analytic continuation to a Euclidean metric. The horizon (co-dimension two surface) splits the system into two sub-systems for which we can define a reduced density matrix $\rho$. 

The corresponding entanglement entropy can be obtained by applying the 
so-called replica method \cite{Callan:1994py, SoloConi, SoloStelle}, which boils down to considering an $n$-fold cover ${E}_n$
of the original (Euclidean) spacetime defined at $n = 1$.  In this case the time coordinate 
$\tau$ is periodic with a period of
$2 \pi n$. Moreover, on the surface $\Sigma$ there is a conical singularity, so that in the small vicinity of $\Sigma$ the total space $E_n$ is locally a direct product $\Sigma \times \mathcal{C}_{n}$, where $ \mathcal{C}_{n}$ is a two-dimensional cone with a deficit angle given by $\delta = 2 \pi (1 - n)$. The entanglement entropy is to be computed on this conifold manifold according to the formula by R\'enyi 
\be
S_n(\rho)=\frac{1}{1-n}\ln {\rm Tr}\rho^n.
\label{renyi}
\ee
Subsequently to get the Von Neumann entanglement entropy a limit $n\to1$ must be taken.
The trace ${\rm Tr} \rho^{n} $ computed for the state described by the density matrix $\rho$ on the conifold $E_n$ has  a natural interpretation as a partition function for the gravitational field configurations over $E_n$.
This construction can be analytically continued to an arbitrary non-integer: $n\rightarrow \alpha$. Therefore, one can define the partition function
\be
Z(\alpha)  = {\rm Tr} \rho^{\alpha } ,
\ee
by the path integral on the field configurations over $E_\alpha$. Defining the quantum effective action as 
\be 
W(\alpha)=-\ln Z(\alpha),
\ee
the entanglement entropy is computed as \cite{Callan:1994py}
\be
S = (\alpha \partial_\alpha -1) W(\alpha)|_{\alpha = 1}.
\label{eq:entropy}
\ee
Using the replica trick the above formula gives the off-shell entanglement entropy \cite{SoloLiving}.

A standard way to evaluate the effective action is to express it as 
\be
W=-\frac12\int^\infty_{\epsilon^2}\frac{ds}s {\rm Tr} K \,, \quad 
{\rm where } \quad 
K=e^{-s\mathcal{D}}
\ee
 is the heat kernel and the operator $\mathcal{D}$ is obtained from the second functional variation of the action (\ref{gravityG}) with respect to gravitational perturbations and will contain both derivatives and curvatures. Above the $\epsilon$ is a UV regulator. We want to consider an expansion of ${\rm Tr}e^{-s\mathcal{D}}$ in which each term 
 has a definite number of derivatives of the metric, 
\begin{equation}
{\rm Tr}e^{-s\mathcal{D}}=\frac1{(4\pi)^{\frac{d}2}}\sum_{m=0} a_m {\cal T}_m (s)\,,
\end{equation}
where ${\cal T}_m (s)$ are homogeneous functions, examples of which appear in \cite{Nesterov:2010yi}. We can thus obtain an expansion in the number of derivatives for both the finite and divergent parts of the quantum effective action. This decomposition is valid both for regular manifolds and manifolds with a conical singularity like $E_\alpha$. If a conical singularity is present, the coefficients $a_m$ can be  decomposed as
\begin{eqnarray}
\hspace{-0.5cm}
a_m(\alpha) = a_m^{\rm reg}(\alpha)+a_m^\Sigma(\alpha) 
= \alpha \, a_m|_{\alpha=1}+a_m^\Sigma(\alpha), 
\label{an}
\end{eqnarray}
where $a_m|_{\alpha=1}$ are the coefficients in the heat kernel expansion on a regular spacetime and $a_m^\Sigma$ are the surface terms given by integrals over the entangling surface $\Sigma$. The surface term for $m=1$ is just the area of the surface $\Sigma$ and it gives 
the area term in the entropy computed by formula \eqref{eq:entropy} and in particular, it will not depend on the terms containing curvatures.
%
The coefficient $a_1^\Sigma(\alpha)$ will, therefore, be determined by the full divergent structure of the quantum effective action, where one should include both bulk terms and additional UV-divergent terms localized on the entangling surface \cite{ordinediretto}. This implies quite complicated running of the area term with the renormalization scale.

Nevertheless, it was suggested in \cite{Cooperman:2013iqr} that one can skip such a straightforward computation by first considering a family of singularity-free spacetimes and afterwards taking a singular conical limit. This procedure allows to compute the entropy of a black hole by just considering the quantum gravitational effective action $W$ on a regular background (RB) and then deforming  the RB to get the effective action $W(\alpha)$ for the $\alpha$-fold covering {$E_\alpha$}. Finally, one applies formula \eqref{eq:entropy} again. Actually, there is a standard procedure relating the curvature terms computed on a smooth manifold {$E$} to the corresponding ones for {$E_\alpha$} \cite{SoloLiving, SoloStelle, SoloRecent, SoloConi}. Therefore we have at our disposal two methods: one of computing the coefficients directly on the singular conifold using heat kernel techniques and the second one of computing them on a RB and eventually taking the singular conical limit. In essence, the two procedures differ by the order of the sequence in which the conical limit is taken: before or after the actual computation of divergent coefficients. We could do this at the beginning or after the resolution of the manifold $E_\alpha$.  In general the two procedures will not produce the same divergent terms, both because of possible contributions from the surface divergences not related to bulk divergences or because of possible non-analytic contributions in $\alpha$. However, the latter should be excluded \cite{SoloRecent} on the basis of analytic continuation used in the definition of R\'enyi entropy. In  \cite{SoloRecent} it was also argued that additional surface divergences can only give contributions at least of order $O((1-\alpha)^2)$, which will therefore drop out of \eqref{eq:entropy}. 
After \cite{SoloConi, SoloStelle} we take into account the contribution of the curvatures induced by the conical singularity and therefore we actually compute the conical entropy.

With the above simplifications for divergent contributions, the running of the area term with the renormalization scale can be completely read out from the bulk UV-divergent effective action. This implies that the term in the entropy which is proportional to the area of the entangling surface $\Sigma$  is determined by the coefficient in front of the Ricci scalar in the effective action (i.e. effective Newton constant). More explicitly
\begin{equation}
S= \frac{A\left(\Sigma\right)}{4 G_{\rm ren}}\,,
\label{eq:ConicalEntropy}
\end{equation}
where $G_{\rm ren}$ is the renormalized value of the Newton constant $G_N$. This is a natural generalization of the Bekenstein-Hawking formula for black hole entropy.

In addition we assume the validity of the general renormalization procedure  described in  \cite{Cooperman:2013iqr}, by treating the dynamics of gravity as the one of a spin-2 field. If the entangling surface is the event horizon of a black hole, then the area term in the renormalized entanglement entropy is the Bekenstein-Hawking entropy and the proportionality factor is given in terms of the renormalized Newton constant as in \eqref{eq:ConicalEntropy}. Furthermore, again following \cite{Cooperman:2013iqr}, the modes on the entangling surface do not actually contribute to divergences of the entanglement entropy neither to the leading nor to the sub-leading terms. In the case of a non-renormalizable quantum theory of gravity (like Einsteinian gravity), this is  given by the resummation of contributions coming from an infinite number of quantum corrections to the Einstein-Hilbert counterterm and this will keep an explicit dependence on the cut-off. 

Let us consider the case in which the full theory consisting of matter and gravity is super-renormalizable
\cite{Tombo, modesto, modestoLeslaw}.
All the quadratic operators in the theory are weakly non-local (or local for standard killers) higher derivative 
operators. In short, the quadratic operator in the graviton fluctuation ${\bf h}$, including the gauge fixing,
reads
\be && \hspace{-0.1cm}
{\bf h} \, {\mathcal H}_{\rm gr} \, {\bf h} 
= 
{\bf h}  \left[ \Box^{\gamma+2}  + \left( \!  a + \sum_i s^{(1)}_i  \! \right) {\cal R} \nabla^{2 \gamma +2} 
\right.
\nonumber \\
&&  \hspace{-0.1cm}
\left. 
+ \left( \! b + \sum_i s^{(2)}_i  \! \right){\cal R}^2 \, \nabla^{2 \gamma }
+ \left( \! c + \sum_i s^{(1)}_i  \! \right) (\nabla{\cal R}) \, \nabla^{2 \gamma +1} \right. \nonumber \\
&&  \hspace{-0.1cm}
\left.
+\left(\! d + \sum_i s^{(1)}_i  \! \right) (\nabla^2{\cal R}) \, \nabla^{2 \gamma}+\ldots\right] \! {\bf h} \,  ,
\ee
where all indices are neglected, while $a,b,c,d,\ldots $ are numerical constants resulting from the variation of the form-factor in its asymptotic limit (\ref{TomboFF}) and of the gauge fixing operator. Finally, 
$s_i^{(1)}, s_i^{(2)}$ are the coefficients in front of the killer operators. 
The reader can refer to the Appendix A for more details about the counterterms in cut-off regularization scheme.

Similar operators will result from taking the second variation of the action with respect to the gauge fields and matter sectors. 

The quantum action for gravity including the contributions of the gauge and matter fields eventually reads
\be
\,\, W_{\rm gr} 
= \frac{1}{2} \log {\rm det} \left( \mathcal{H}_{\rm gr} \right) 
\propto B_0 \, k^4 + B_2 \, k^2 + B_4 \, \log \left( \frac{k^2}{\mu^2} \right) \! , \nonumber
\label{divB}
\ee
where $k$ is the UV cut-off and $\mu$ is the typical scale of renormalization.
In dimensional regularization there is no contribution to $B_0$ and $B_2$, and therefore there is no renormalization of $G_N$. 
In cut-off regularization scheme the coefficients $B_2$ and $B_4$ will depend linearly on the killers' 
coefficients $s_i^{(1)}, s_i^{(2)}$ respectively. Therefore, we can always get vanishing divergent contributions to the $G_N$ and the coupling constants multiplying $R^2$ or ${\bf Ric}^2$ beta functions by tuning the values of these $s_i^{(1)}$ and $ s_i^{(2)}$. $B_0$, $B_2$ and $B_4$ do not depend on the gauge fixing parameters \cite{shapiro1} so that this result is completely gauge-independent in whatsoever  regularization scheme. Actually, it was crucial in \cite{SoloLiving, SoloRecent} to use the cut-off regularization scheme to properly compute the entanglement entropy, however, here we give results also in DR for completeness, since we use methods based on effective gravitational action $W(\alpha)$. We also add that the divergences as displayed in \eqref{divB} are the only ones that we encounter, even if we consider our theory on a manifold with conical singularity, due to the background-independence (on a smooth manifold) of the UV-divergences.

In order to calculate the associated conical entropy, the effective action should be evaluated on the regularized manifold ${E}_\alpha \equiv \Sigma \times \mathcal{C}_{\alpha}$ and the singular limit of conifold should be taken afterwards. 
Since we do not have any renormalization of the Newton constant we also do not have divergent contributions
to the entropy proportional to the area. However, due to the classical 
Einstein-Hilbert term, we still have the usual finite leading contribution $A/(4G_N)$.
Moreover, we will have other  finite contributions to the entropy due to local and non-local finite quantum corrections to the effective action. 
This outcome does not change when matter without self-interactions or a weakly non-local matter (or gauge theory) is coupled to gravity. 


If the theory (eventually including also the gauge fields (\ref{gauge}) and matter) is UV-finite, we get the remarkable result that the leading contribution to the entropy is the finite one coming from the classical Einstein-Hilbert term. It has been noticed that in general the conical entropy \eqref{eq:ConicalEntropy} is not positive-definite and is gauge and renormalization scheme dependent. For the proposed UV-finite theory of gravity we can solve all these drawbacks because we do not have RG running of $G_N$.

Now we take a closer look at the coupling of matter to the purely gravitational theory \eqref{theory}. In particular, the case of matter fields with non-minimal coupling to gravity has risen some puzzles \cite{Fursaev:1994pq,Larsen:1995ax,SoloRecent} as to what the correct procedure to compute the entanglement entropy is. 

The problems seem to arise from the wish to retain the interpretation of renormalized entanglement entropy as a state-counting. This statistical interpretation is quite natural in the case of physical regulators, like cut-off by a UV scale e.g., but  when gravity is involved, covariant regulators, such as Pauli-Villars 
\cite{Diaz:1989nx, Codello:2015oqa} and the heat kernel regularization, should be preferred and there is no obvious way of carrying out such a counting. This has led to attempts to distinguish statistical and conical definitions of entropy, arguing that the latter is marred by such unphysical features as not being positive definite and being gauge- and regulator-dependent \cite{SoloRecent}. On the other hand, the idea that it can be the more sensible choice in the presence of gravity has been supported \cite{Cooperman:2013iqr} on the basis of the fact that the lack of a statistical interpretation is a common feature of models with the UV-divergent part in covariant regulators. 

Let us now consider a theory in which gravity and gauge interactions are weakly non-local whereas 
fermionic and scalar sectors are local, just as in the standard model of particle physics.
As explained in \cite{SoloLiving, SoloRecent}, the renormalization of the Newton constant due to the fermionic and scalar matter is such that the entanglement entropy coincides with the Bekenstein-Hawking entropy. In our case, due to the absence of divergent contributions to $G_N$ coming from the gauge and gravitational sectors
in DR, or in cut-off regularization when suitable killers are included (\ref{CoffK}), we arrive at the same conclusion for the conical entropy. We want to stress that also in this case the conical entropy, even if UV-divergent, is positive-definite and gauge-independent as a consequence of only finite contributions coming from the gravity and gauge sector.

If  we could switch off the gravitational and gauge interactions 
(no bare Newton constant is present in the theory) 
the whole entanglement entropy would be given by  the ``universally divergent" 
contribution computed in section 7.1 of \cite{SoloRecent}.
%
As stated in 
\cite{SoloRecent} this could provide a natural explanation of the statistical origin of the black hole entropy.
On the other hand, if $G_N$ is in the bare action, but it is not renormalized by gravitons and gauge bosons
for the reasons just explained, the above 
interpretation of the Bekenstein-Hawking entropy is still valid and has a universal character. 
Indeed, 
the non-renormalization of $G_N$ by gravitons and gauge bosons is one-loop exact 
because for $L > 1$ internal gravitational and gauge boson lines make every loop diagram convergent.

We conclude with the following statement: 
{\em  only matter participates in giving a ``universal" renormalization to both
the Einstein-Hilbert term and the conical entropy. }

Let us remark that in non-super-renormalizable theories, and in particular in two-derivative theories,
$G_N$ (in the cut-off scheme) gets renormalized at any order in the loop expansion and the above interpretation of the black hole entropy is likely lost.  This is in particular the case for the theory of a scalar field conformally coupled to gravity, where only the one-loop correction to the gravitational constant vanishes whereas higher loop divergent terms are expected. 
Only in super-renormalizable theories the interpretation given in \cite{SoloLiving, SoloRecent} has a universal character independent on the perturbative order.

We summarize that for a super-renormalizable theory we only found one-loop divergences and the dependence on the cut-off disappears by a one-loop exact (for the minimal super-renormalizable theory) 
renormalization of a finite number of couplings. 
Actually, for a finite quantum field theory of gravity no renormalization of the Newton constant is needed. It is believed that in a complete theory of quantum gravity there is a fundamental length that can be physically probed. If we associate the usual UV divergence of the entanglement entropy with the presence of correlated modes with arbitrarily short wavelengths, it is natural to expect a finite entanglement entropy for the theories just discussed. We found that in our theories the conical entropy is finite without explicitly introducing any cut-off or regulator scale, which could correspond to such a
fundamental length. We emphasize that our results were obtained in continuous field theory. Actually, we explicitly found that the conical entropy of black holes is finite 
in a consistent theory of gravity coupled to matter as a mere consequence 
of the finiteness of the fundamental theory, which we reviewed in section II. Therefore, the leading area law contribution to the entanglement entropy evaluated with the replica trick (see \cite{SoloLiving} and references within) coincides with the analogue entropy in Einstein-Hilbert classical gravity.

Our result is only based on the presence of an event horizon independently on the exact or approximate nature of the solution. Our analysis cannot be applied to compact objects without an event horizon \cite{KSno}. 
Therefore, once ascertained that the theory allows some kind of black hole solutions, we can apply the analysis developed in this section, where it was proved that the finiteness of the theory, in DR or in cut-off scheme, implies that also the conical entropy is finite. 

Therefore, we have shown that in finite non-local quantum gravity we are able to overcome the long standing tension between always convergent classical Wald entropy and Entanglement entropy, which is usually divergent in quantum field theory.

\section{Classical and quantum Wald entropy}
In this section we want to discuss the relationship between the conical entropy that we computed in the previous section and  the Wald entropy. 
The formula \eqref{eq:entropy} for the conical
entropy can be rewritten as
\begin{equation}
S=2\pi \int_\Sigma \frac{\partial \mathcal{L}}{\partial R^{\alpha\beta}{}_{\mu\nu}}\epsilon_{\mu\nu}\epsilon^{\alpha\beta}\,,
\label{eq:Wald}
\end{equation}
which is exactly the Wald entropy \cite{Wald:1993nt,Iyer:1994ys}. Above $\epsilon$'s denote completely antisymmetric Levi-Civita tensors on two-dimensional spacetimes, respectively on a disk $D_2$ and the horizon surface $\Sigma$. We notice that Wald's Noether charge method is on-shell so that the metric in the expression for the Wald entropy is supposed to satisfy the gravitational field equations. On the contrary, the conical singularity method is an off-shell method valid for any metric that describes a black hole horizon \cite{SoloLiving}. We believe that the identification of the conical entropy \eqref{eq:entropy} with the Wald entropy \eqref{eq:Wald} supports even more the fact that the definition of the entropy presented in section IV is a physically meaningful one.

Using the results of the previous section we can infer that for finite gravitational theories the leading area law term of the Wald entropy does not differ at quantum level from its classical counterpart. However, the full quantum effective action, including the finite quantum contributions, will give corrections to the classical Wald
entropy. The Wald entropy formula can be applied to the classical as well as to the quantum effective action because it is defined for any action functional.  When we will use the Wald formula \eqref{eq:Wald} for the quantum effective action, then we will call the related entropy quantum entropy. 
Therefore, in a quantum effective action that has only finite contributions, we can compute the finite contributions to the Wald entropy simply using formula \eqref{eq:Wald}, where we treat the effective action as a classical one.
In the case of a simple spherically symmetric metric of the type
\begin{equation}
ds^2=-f(r) dt^2 + f(r)^{-1}dr^2+r^2 d\Omega^2,
\end{equation}
the Wald entropy can be recast as a closed integral over a cross section of the horizon. For the classical action \eqref{gravityG} with  $ V_K=0$, the following general formula can be derived 
\be
\hspace{-0.5cm}S_{\rm W} = \frac{A}{4 G_N} 
\left[ 1+ \left( 2 \gamma_0(\Box) +  \gamma_2(\Box)  +  2\gamma_4(\Box) \right) R 
\right]_{r_H} \! ,
\label{eq:NLWald}
\ee
where the label ${r_H}$ stands for: evaluated at the event horizon. 
For the sake of simplicity we here omitted two other contributions 
that can be found in \cite{Myung}. 
Formula (\ref{eq:NLWald}) can actually be rewritten as
\be
S_{\rm W} = \frac{A}{4 G_N} 
\left[ 1+ \left( 2 \gamma^\prime_0(\Box) +  \gamma^\prime_2(\Box) \right) R 
\right]_{r_H}  \,,
\ee
where we used the following basis for the operators in the action
\begin{eqnarray}
&& \mathcal{L}_{\rm g} = -  2 \kappa_{D}^{-2} \, \sqrt{|g|} 
\left[ { R} 
+
{R} \, 
 \gamma^\prime_0(\Box)
 {R} 
 + {\bf Ric} \, 
\gamma^\prime_2(\Box)
 {\bf Ric}  
 \right. 
 \nonumber 
 \\
 && \left.
 \hspace{2.5cm}
 + {\rm GB}_ {\gamma^\prime_4(\Box)}
+ {V_K} \, 
\right],
\label{gravityGprime}
\end{eqnarray}
and we introduced the non-local generalization of a Gauss-Bonnet density, namely 
\be
&& \hspace{-1cm} 
{\rm GB}_ {\gamma^\prime_4(\Box)}={\bf Riem}\,\gamma^\prime_4(\Box){\bf Riem}-4{\bf Ric}\,
\gamma^\prime_4(\Box){\bf Ric}
\nonumber \\ 
&& \hspace{0.7cm}
+{ R}\,\gamma^\prime_4(\Box){R} \, , 
\ee
and $\gamma^\prime_0=\gamma_0-\gamma_4$, $\gamma^\prime_2=\gamma_2+4\gamma_4$, 
$\gamma^\prime_4=\gamma_4$. 
In this basis the partial entropy (\ref{eq:NLWald}) does not depend on $ \gamma^\prime_4$, which happens to be exactly  the form-factor not appearing in the expression for the propagator \eqref{propagator} on a flat spacetime. However, the other contributions in \cite{Myung} still depend on $\gamma_4^\prime$.
 %


For the most general theory \eqref{theory} compatible with unitarity the Wald entropy 
in $D=4$ is: 
\be
\hspace{-0.5cm}
S_{\rm W, nl}^{D=4} = \frac{A}{4 G_N} 
\left[ 1+ \left( 2 \gamma _4-\frac{ \left(e^{H_0}-e^{H_2}\right)}{3 \Box} \right) R \right]_{r_H} \!\!\!\!\!\! .
\ee

Finally, at quantum level the form-factors receive corrections strictly related to the quantum properties of the theory.
For a super-renormalizable theory, logarithmic quantum corrections appear and the Wald quantum entropy (labeled by Wq) reads:  
\be
&& \hspace{-0.1cm}
S_{\rm Wq} = \frac{A}{4 G_N} 
\Big\{ 1+ \Big[ 2 \gamma_0(\Box) +{ 2} \beta_0 \log \left( \frac{- \Box}{\mu^2} \right) +  \gamma_2(\Box) \nonumber\\
&&\hspace{-0.1cm}
+ \beta_2 \log \left(  \frac{ - \Box}{\mu^2} \right) +  2\gamma_4(\Box)
+ { 2} \beta_4 \log \left( \frac{- \Box}{\mu^2} \right)   \Big] R + \dots 
\Big\}_{r_H} \!\!\! , \nonumber 
\ee
where the beta functions are rescaled by the Newton constant $G_N$. 
In particular in the local Stelle theory $\gamma_0(\Box)$, $\gamma_2(\Box)$, and $\gamma_4(\Box)$ are just constants. 

\section{Conclusions}
In this paper we explicitly showed 
that in a 
polynomial or quasi-polynomial (ghost-free) higher derivative (or weakly non-local)
gravitational theory coupled to matter 
the conical entropy  for a black hole horizon, due to classical terms and bulk divergences, is finite and  coincides with the area term of  Wald entropy. We emphasize that any super-renormalizable theory can be made finite according to the procedure described in section III. 
The matter sector is also properly chosen to be quasi-polynomial (ghost-free) or weakly non-local in order to have a 
super-renormalizable action for all fundamental interactions. 
Quasi-polynomiality, or weak non-locality, is crucial to achieve 
unitarity and super-renormalizability at the same time. 
In dimensional regularization an appropriate higher derivative kinetic operator is sufficient to make 
the beta function of the Newton coupling zero. In the cut-off regularization scheme (with or without using heat-kernel technique) 
the addition of one extra vertex interaction, which is cubic in the curvature, is sufficient to make the beta function for the Einstein-Hilbert operator vanishing. This is an explicit example of a theory in which interactions do matter and make the difference with respect to the results obtained for free theories \cite{SoloLiving}. We emphasize that in the paper we followed the strategy of reading the entanglement entropy from the effective action and this method easily tells us a lot about the UV divergences of the latter.

Moreover, when weakly non-local gravitational and gauge interactions are coupled to the usual local action for standard matter (scalars and fermions), the gravitational constant has a universal renormalization due to the matter content only. This theory is not super-renormalizable anymore, but it is not affected by
the interpretational problems of conical entropy that may be found in the literature \cite{SoloRecent}.  

Finally, we evaluated the Wald entropy for a wide class of local and non-local classical and quantum 
actions finding agreement with the conical entropy for the terms proportional to the area. We also considered contributions from higher derivative terms, where form-factors show up explicitly. This is a further confirmation of the physical relevance of the conical entropy of black holes whose computation has been performed in this paper. It has been observed \cite{Cooperman:2013iqr,ordinediretto} that in the case of gravitational fluctuations this procedure may miss some contributions as a consequence of the fact  that the regulated metric does not satisfy the vacuum Einstein equations, which seems to point at additional dynamical degrees of freedom inconsistent with a theory of pure gravity. Whereas to discuss this point in detail is beyond the scope of this paper, we notice that it has been recently argued \cite{SoloRecent} that such additional gravitational modes appearing on the singular manifold are an artefact of the orbifold definition and should be excluded upon considering the $n$-fold cover, which is required by the replica trick. So the orbifold and the $n$-fold cover are not in general analytically related to each other and the latter supports just the gravitational modes on a regular background. We think  this argument, even in the absence of a more physical mechanism to exclude non-analytical contributions, should sufficiently support the relevance of the computation presented in this note.

 It is also conceivable that the surface divergences that we did not take into account may actually give no contribution 
in the context of a theory of gravity finite in the bulk. In fact, in such a finite theory of gravity like the one discussed here, the dependence on the regulator of the conical singularity should disappear once the fundamental scales of the theory are introduced and so no additional massless degrees of freedom should appear. On the other hand, if present, such contributions could also be cancelled by switching on appropriate operators localized on the entangling surface. In both cases, the results presented here would then become relevant for the full renormalized 
entropy computed through the Callan-Wilczek formula. 

We decided to consider the contribution of the curvatures in non-minimal couplings due to the conical singularity even in flat spacetime. This means that we computed the conical entropy and proved that for super-renormalizable or finite theories this quantity is positive-definite and gauge-independent making it a good quantity for black holes' entropy.  
We are aware that this may be problematic for theories on flat spacetimes. As far as we know all the efforts there with Rindler horizon and Rindler observers are quite unsuccessful and the entanglement for such horizon is always divergent. 
Here we did not attempt to solve this problem, but we only concentrated on gravity and 
black holes' horizons, for which our choice looks very natural and consistent with general covariance. 
However, we believe that on flat spacetime the killing of the beta function for $G_N$ should actually work the same because the divergences are independent on the background. We here added a killer term that is a vertex on flat spacetime and hence it does not vanish there. 

In this paper we showed that a gravitational theory is finite if ``very interacting". 
Nevertheless, if there are no interactions (in particular non-minimal ones), the conical as well as the entanglement
entropy turns out to be divergent again.
Indeed, in our work we pointed out
that interactions do matter and the operators needed to achieve super-renormalizability are not sufficient. We need more interaction terms
that are named killers in this paper. 
The entanglement entropy in flat spacetime is divergent because essentially based on a free theory, but a theory with proper interactions should overcome this issue as suggested by several string theory computations  \cite{Dabholkar:1994gg, Dabholkar:1994ai, Dabholkar:2001if, He:2014gva,Hartnoll:2015fca}. In the spirit of \cite{Paddy}, we found that for the specific class of theories described above the conical entropy can correctly account for the expected contribution of interactions.

Once more we would like to remark that 
the goal of this project was not to find a microscopic origin for the black hole entropy, but to point out that 
the conical entropy is finite in a UV-complete theory. 
Moreover, in this class of theories we were able to remove the tension between the finite Wald Entropy and the quantum entropy, which is generically divergent in quantum field theory. 
Regarding the statistical interpretation both the Wald entropy and the conical entropy, which are the only ones used in the paper, do not have any statistical meaning. 
Any further interpretation is beyond the scope of this paper. 

All the results obtained in this paper can be easily exported to Lee-Wick gravitational theories \cite{shapiro1, shapiro2, shapiro3, lm5} by just replacing the non-local form-factors with appropriate polynomials \cite{lm5}. 


\section*{Acknowledgements}
We would like to thank Ling-Yan Hung and Arpan Bhattacharya for relevant advice and discussions on several important issues in entanglement entropy during the development of this work. We would also like to thank Daniel Litim for suggestions about the importance of cut-off regularization in higher derivative theories.
%


\begin{widetext}

\section*{Appendix A: general divergent one-loop integrals in cut-off regularization scheme} 
The main divergent one-loop integral in a $D$-dimensional spacetime for an asymptotically monomial higher derivative theory reads: 
\be 
\int \!\! \frac{d^D q}{(2 \pi)^D}  \left\{ \prod_{i=1}^{s} 
\frac{1}{[(q+p_i)^2 ]^{n}} \right\} \! P(q)_{2 s  n}  . 
\label{integralCutOff0}
\ee
$P(q)_{2sn}$ is a polynomial function of degree $2sn$ in the integration momentum $q$ 
(generally it also relies on the external momenta $\bar{p}_a$), 
$p_i = \sum_{a=1}^i \bar{p}_a$.
The positive integer $n$ is: $n = \gamma +{\rm N} +2$ for the graviton $h_{\mu\nu}$, while it is 
respectively 
$n = \gamma +{\rm N} +1$ and $n=1$ for the ghosts $b_{\mu}$ and $C,\bar{C}$ (see \cite{modestoLeslaw} for more details about the action for ghosts). Finally, $s$ is the number of external legs at one loop. 
Once more we would stress that the computation below is one-loop exact because as showed in the main text there are 
no divergences from two loop onwards. 
We can write, as usual,

\be
&& 
\prod_{i=1}^{s} 
\frac{1}{[
(q+p_i)^2 + M^2]^{n}}  
 = {\rm c} \!  \int_0^1 \! \left( \prod_{i=1}^s x_i^{n-1} d x_i \!  \right) 
\delta\left( 1 - {\sum_{i=1}^s x_i} \right)
 \frac{1}{[q^{\prime 2} + M^{2}]^{sn}}  \, , \quad {\rm where}\label{withqprime}\\
 &&  
 q^{\prime} = q + \! \sum_{i=1}^s x_i p_i  \, , \quad  
 M^{2} =  - \left( \sum_{i=1}^s x_i p_i \right)^{\!\! 2} \!\! + 
 \sum_{i=1}^s p_i^2 x_i  \, , \nonumber  
\ee
where ${\rm c} = {\rm constant}$. 
In (\ref{withqprime}), we move outside the convergent integrals in $x_i$ and we replace 
$q^{\prime}$ with $q$ again. The outcome reads 
\be
\int \!\! \frac{d^D q}{(2 \pi)^D} 
\frac{P^{\prime}(q, p_i, x_i)_{2 s  n} }{( q^2 + M^{2})^{s n }} \, .
\label{int2b}
\ee
Using Lorentz invariance and neglecting the argument $x_i$, we replace the polynomial 
$P^{\prime}(q, p_i, x_i)_{2 s n}$ with a polynomial of degree $s \times n$ in $q^2$, 
namely $P^{\prime \prime}(q^2, p_i)_{s n}$. {In cut-off regularization scheme we have to integrate (\ref{int2b}) up to a cut-off scale $\Lambda_{\rm c}$}
\be
\int_0^{\Lambda_{\rm c} } \!\!\!  \frac{d^D q}{(2 \pi)^D}
\frac{P^{\prime \prime}(q^2, p_i)_{ s n} }{( q^2 + M^{2} )^{ s n }} \, .
\label{int3b}
\ee
We can decompose the polynomial $P^{\prime \prime}(q^2, p_i)_{ s n} $ 
in a product of external and internal momenta
only to obtain the divergent contributions. Below we consider only parts of this polynomial which give contributions to divergences, namely
\be
P^{\prime \prime}(q^2, p_i)_{s n} = \sum_{\ell=0}^{\left\lfloor  D/2 \right\rfloor} \alpha_{\ell}(p_i) q^{2 s n-2 \ell}  
= \alpha_0 q^{2 s n}  + \alpha_1(p_i) q^{2s  n-2}  + \alpha_2(p_i)    q^{2 s n - 4 } 
+ \ldots  + \alpha_{\left\lfloor  D/2 \right\rfloor}(p_i) q^{2s  n-2\left\lfloor  D/2 \right\rfloor}. 
\ee
By changing variables to $y = |q|^2/M^{2}$ 
the integral (\ref{int3b}) for the case of even dimension $D$ turns  into: 
{
\be
 && 
\int_0^{\frac{\Lambda^2_{\rm c}}{M^{2}}   } \! d y \,
y^{\frac{D-2}{2}}
\frac{M^{D}}{(1+y)^{s n}}
\Big[  \alpha_0 \, y^{s n}  + 
\frac{\alpha_1(p_i)}{ M^{2} } \, y^{sn -1} + 
\frac{\alpha_2(p_i)}{ M^{4} } \, y^{sn -2}
+ 
\frac{\alpha_3(p_i)}{ M^{6} } \, y^{sn -3}
 + \dots 
\Big]  \nonumber   \\
&& 
= 
\int_0^{\frac{\Lambda^2_{\rm c}}{M^{2}}   } \! d y \,
y^{\frac{D-2}{2}}
\frac{M^{D}}{(1+y)^{sn}}
   \! \sum_{\ell = 0}^{  D/2} \frac{\alpha_{\ell} (p_i)}{ M^{2 \ell } } y^{s n - \ell}  
   \nonumber \\
   &&  
= 
   \alpha_0 \Bigg[ a_D^{(0)} 
 \Lambda_{\rm c}^D  
  + a_{D-2}^{(0)} M^2   \Lambda_{\rm c}^{D-2} 
  +\ldots + a_0^{(0)} M^D 
  \log \left(\frac{\Lambda_{\rm c}   }{  M }\right)^{\!\! 2} 
  \Bigg]
  \nonumber \\ 
  &&  \hspace{0cm}
  +
  \alpha_1(p_i) \Bigg[ 
   a_{D-2}^{(1)} \Lambda_{\rm c}^{D-2} 
  + a_{D-4}^{(1)} M^2 \Lambda_{\rm c}^{D-4}  
   + \ldots + a_0^{(1)} M^{D-2}
  \log \left(\frac{\Lambda_{\rm c}   }{  M }\right)^{\!\! 2} 
  \Bigg] 
 %
  + \ldots  +  \nonumber \\
  && +
  \alpha_{  D/2 }(p_i)  \,a_0^{\left(  D/2 \right)}
   \log \left(\frac{\Lambda_{\rm c}   }{  M }\right)^{\!\! 2}  
     \nonumber  
     \ee
     \be
&& = 
   \underbrace{ \alpha_0 \, a_D^{(0)} 
 \Lambda_{\rm c}^D }_{\sim\,\, k^D \,  \sqrt{|g|} }
  +  \underbrace{ \alpha_0 \, 
  a_{D-2}^{(0)} M^2   \Lambda_{\rm c}^{D-2} }_{\sim \,\, k^{D-2} \, R }
  +\ldots 
  + \underbrace{ \alpha_0 \, a_0^{(0)} M^D 
  \log \left(\frac{\Lambda_{\rm c}   }{  M }\right)^{\!\! 2} }_{\sim \,\, \log k^2 \, R^{D/2} }
  \nonumber \\&&  \hspace{0cm}
  +
  \underbrace{
  \alpha_1(p_i) \, 
   a_{D-2}^{(1)} \Lambda_{\rm c}^{D-2} }_{\sim \,\, k^{D-2} \, R }
  +  \underbrace{  \alpha_1(p_i) \, 
  a_{D-4}^{(1)} M^2 \Lambda_{\rm c}^{D-4}}_{\sim \,\, k^{D-4} \,  R^2 } 
   + 
   \ldots + 
   \underbrace{
    \alpha_1(p_i) \, 
   a_0^{(1)} M^{D-2}
  \log \left(\frac{\Lambda_{\rm c}   }{  M }\right)^{\!\! 2}  }_{\sim \,\, \log k^2 \, R^{D/2} }
     \nonumber \\
 &&  \hspace{0cm}
  + \ldots  + 
 \underbrace{  \alpha_{  D/2 }(p_i) \, a_1^{\left(  D/2 \right)}
     \log \left(\frac{\Lambda_{\rm c}   }{  M }\right)^{\!\! 2} }_{ \sim \,\, \log k^2 \, R^{D/2}  }
\ee
The corresponding covariant structure of divergences is explicitly displayed under braces. 
} 

We now rename $\Lambda_{\rm c} \equiv k$ and explicitly show the divergent contributions in the effective action
for $D=4$,
\be
\Gamma^{(1)}_{k \, {\rm div} } =  
 \!\int\! d^4 x \sqrt{|g|} \left[ - \frac{ \beta_{G_N} }{2} k^2 R 
+ \frac{\beta_{\bar{\lambda}}}{4} k^4  + \beta_{R^2} \log k^2 \, R^2 + \beta_{{\rm Ric}^2} \log k^2 \, R_{\mu\nu} R^{\mu\nu}
\right] \, .
\ee

\end{widetext}


\begin{thebibliography}{99}




\bibitem{Strominger:1996sh} 
  A.~Strominger and C.~Vafa,
  Phys.\ Lett.\ B {\bf 379}, 99 (1996)
  [hep-th/9601029].
  
\bibitem{Susskind:1994sm} 
  L.~Susskind and J.~Uglum,
  Phys.\ Rev.\ D {\bf 50}, 2700 (1994)
  [hep-th/9401070].
  
  
  \bibitem{SoloLiving}
  S.~N.~Solodukhin,
  Living Rev.\ Rel.\  {\bf 14}, 8 (2011)
  [arXiv:1104.3712 [hep-th]].
  

 \bibitem{SoloConi} 
  S.~N.~Solodukhin,
  Phys.\ Rev.\ D {\bf 51}, 609 (1995)
  [hep-th/9407001].
  
\bibitem{SoloStelle} 
  D.~V.~Fursaev and S.~N.~Solodukhin,
  Phys.\ Rev.\ D {\bf 52}, 2133 (1995)
  [hep-th/9501127].

\bibitem{Larsen:1995ax} 
  F.~Larsen and F.~Wilczek,
  Nucl.\ Phys.\ B {\bf 458}, 249 (1996)
  [hep-th/9506066].
  
\bibitem{Callan:1994py} 
  C.~G.~Callan, Jr. and F.~Wilczek,
  Phys.\ Lett.\ B {\bf 333}, 55 (1994)
  [hep-th/9401072].

  \bibitem{SoloRecent}
  S.~N.~Solodukhin,
  Phys.\ Rev.\ D {\bf 91}, no. 8, 084028 (2015)
  [arXiv:1502.03758 [hep-th]].
  
  

  
%
%
%
%
%

%
%
%



  
  
  
  
  
  
  
  
  
 
  
    
  \bibitem{Krasnikov}
  N.~V.~Krasnikov,
  Theor.\ Math.\ Phys.\  {\bf 73}, 1184 (1987)
  [Teor.\ Mat.\ Fiz.\  {\bf 73}, 235 (1987)].


\bibitem{Tombo} 
E. T. Tomboulis
[hep-th/9702146v1].

\bibitem{Khoury}
  J.~Khoury,
  Phys.\ Rev.\ D {\bf 76}, 123513 (2007)
  [hep-th/0612052].


\bibitem{modesto}
  L.~Modesto,
  Phys. \ Rev. \ D {\bf 86}, 044005 (2012)
  [arXiv:1107.2403 [hep-th]]; 
  L.~Modesto,
  Astron. Rev. 8.2 (2013) 4-33
  [arXiv:1202.3151 [hep-th]].
  
  
  
  
  
  \bibitem{modestoLeslaw}
  L.~Modesto and L.~Rachwal,
  Nucl.\ Phys.\ B {\bf 889}, 228 (2014) 
  [arXiv:1407.8036 [hep-th]].
\bibitem{universality} 
  L.~Modesto and L.~Rachwal,
  Nucl.\ Phys.\ B {\bf 900}, 147 (2015) 
  [arXiv:1503.00261 [hep-th]].
  L.~Modesto, M.~Piva and L.~Rachwal, 
Phys.\ Rev.\ D {\bf 94}, no. 2, 025021 (2016) [arXiv:1506.06227 [hep-th]].   
  
 


\bibitem{Briscese:2013lna}
  F.~Briscese, L.~Modesto and S.~Tsujikawa,
  Phys.\ Rev.\ D {\bf 89}, 024029 (2014)
  [arXiv:1308.1413 [hep-th]];
    A.~S.~Koshelev, L.~Modesto, L.~Rachwal and A.~A.~Starobinsky, JHEP {\bf 1611}, 067 (2016)  [arXiv:1604.03127 [hep-th]]

\bibitem{Dona} 
  P.~Don\`a, S.~Giaccari, L.~Modesto, L.~Rachwal and Y.~Zhu,
  JHEP {\bf 1508}, 038 (2015) 
  [arXiv:1506.04589 [hep-th]].

\bibitem{Cnl1}
  G.~Calcagni, M.~Montobbio and G.~Nardelli,
  Phys.\ Lett.\ B {\bf 662}, 285 (2008)
  [arXiv:0712.2237 [hep-th]];
  G.~Calcagni and G.~Nardelli,
  Phys.\ Rev.\ D {\bf 82}, 123518 (2010)
  [arXiv:1004.5144 [hep-th]].


\bibitem{Mtheory}
  G.~Calcagni and L.~Modesto,
  Phys.\ Rev.\ D {\bf 91}, no. 12, 124059 (2015)
  [arXiv:1404.2137 [hep-th]].
  




\bibitem{Modesto:2013jea}
  L.~Modesto and S.~Tsujikawa,
  Phys.\ Lett.\ B {\bf 727}, 48 (2013)
  [arXiv:1307.6968 [hep-th]];
  
  
  
  
  \bibitem{shapiro1}
  M.~Asorey, J.~L.~Lopez and I.~L.~Shapiro,
  Int.\ J.\ Mod.\ Phys.\ A {\bf 12}, 5711 (1997)
  [hep-th/9610006].


\bibitem{Cooperman:2013iqr} 
  J.~H.~Cooperman and M.~A.~Luty,
  JHEP {\bf 1412}, 045 (2014)
  [arXiv:1302.1878 [hep-th]].



  
\bibitem{Dabholkar:1994gg} 
  A.~Dabholkar,
  Phys.\ Lett.\ B {\bf 347}, 222 (1995)
  [hep-th/9409158].
  
\bibitem{Dabholkar:1994ai} 
  A.~Dabholkar,
  Nucl.\ Phys.\ B {\bf 439}, 650 (1995)
  [hep-th/9408098].
  
\bibitem{Dabholkar:2001if} 
  A.~Dabholkar,
  Phys.\ Rev.\ Lett.\  {\bf 88}, 091301 (2002)
  [hep-th/0111004].
  
\bibitem{He:2014gva}
  S.~He, T.~Numasawa, T.~Takayanagi and K.~Watanabe,
  JHEP {\bf 1505} (2015) 106
  [arXiv:1412.5606 [hep-th]].
\bibitem{Hartnoll:2015fca} 
  S.~A.~Hartnoll and E.~Mazenc,
  Phys.\ Rev.\ Lett.\  {\bf 115}, no. 12, 121602 (2015)
  [arXiv:1504.07985 [hep-th]].
  
  


  \bibitem{Paddy}
  T.~Padmanabhan,
  Phys.\ Rev.\ D {\bf 82}, 124025 (2010)
  [arXiv:1007.5066 [gr-qc]].

  
\bibitem{Hung:2011xb} 
  L.~Y.~Hung, R.~C.~Myers and M.~Smolkin,
  JHEP {\bf 1104}, 025 (2011)
  [arXiv:1101.5813 [hep-th]].
  
  
  
\bibitem{Casini:2014yca} 
  H.~Casini, F.~D.~Mazzitelli and E.~Test\'e,
  Phys.\ Rev.\ D {\bf 91}, no. 10, 104035 (2015)
  [arXiv:1412.6522 [hep-th]].
  


\bibitem{ModestoMoffatNico}
L. Modesto, J. W. Moffat, P. Nicolini,
Phys. Lett. B 695, 397-400 (2011)
[arXiv:1010.0680 [gr-qc]].


\bibitem{Frolov:2015bta} 
  V.~P.~Frolov,
  Phys.\ Rev.\ Lett.\  {\bf 115}, no. 5, 051102 (2015)
  [arXiv:1505.00492 [hep-th]];
%
  V.~P.~Frolov, A.~Zelnikov and T.~de Paula Netto,
  JHEP {\bf 1506}, 107 (2015)
  [arXiv:1504.00412 [hep-th]]; 
%
  V.~P.~Frolov and A.~Zelnikov,
  arXiv:1509.03336 [hep-th].



 \bibitem{BambiMalaModesto2}
  C.~Bambi, D.~Malafarina and L.~Modesto,
  Phys.\ Rev.\ D {\bf 88}, 044009 (2013)
  [arXiv:1305.4790 [gr-qc]].
  
\bibitem{BambiMalaModesto} 
  C.~Bambi, D.~Malafarina and L.~Modesto,
  Eur.\ Phys.\ J.\ C {\bf 74}, 2767 (2014)
  [arXiv:1306.1668 [gr-qc]].


\bibitem{calcagnimodesto} 
  G.~Calcagni, L.~Modesto and P.~Nicolini,
  Eur.\ Phys.\ J.\ C in press
  [arXiv:1306.5332 [gr-qc]].
  



\bibitem{koshe1}
  A.~S.~Koshelev,
  Class.\ Quant.\ Grav.\  {\bf 30}, 155001 (2013)
  [arXiv:1302.2140 [astro-ph.CO]]; 
%
 A.~S.~Koshelev and S.~Y.~Vernov,
  Phys.\ Part.\ Nucl.\  {\bf 43}, 666 (2012)
 [arXiv:1202.1289 [hep-th]].
%
  %
  A.~S.~Koshelev,
  Rom.\ J.\ Phys.\  {\bf 57}, 894 (2012)
  [arXiv:1112.6410 [hep-th]].
%
  S.~Y.~Vernov,
  Phys.\ Part.\ Nucl.\  {\bf 43} (2012) 694
  [arXiv:1202.1172 [astro-ph.CO]]. 
%
  A.~S.~Koshelev and S.~Y.~Vernov,
  arXiv:1406.5887 [gr-qc].
  
  
  

\bibitem{Buoninfante:2018xiw} 
  L.~Buoninfante, A.~S.~Koshelev, G.~Lambiase and A.~Mazumdar,
  arXiv:1802.00399 [gr-qc].


  
\bibitem{Koshelev:2018hpt} 
  A.~Koshelev, J.~Marto and A.~Mazundar,
  arXiv:1803.00309 [gr-qc].





\bibitem{exactsol} 
 Yao-Dong Li, L.~Modesto and L.~Rachwal,
  JHEP {\bf 1512}, 173 (2015)
  [arXiv:1506.08619 [hep-th]].

\bibitem{BrisceseClassical} 
  F.~Briscese and M.~L.~Pucheu,
  arXiv:1511.03578 [gr-qc].
  
    \bibitem{appear} G. Calcagni, A. Koshelev, S. Kumar, L. Modesto, L. Rachwal,  to appear.
  

\bibitem{Wald:1993nt} 
  R.~M.~Wald,
  Phys.\ Rev.\ D {\bf 48}, 3427 (1993)
  [gr-qc/9307038].
  
\bibitem{Iyer:1994ys} 
  V.~Iyer and R.~M.~Wald,
  Phys.\ Rev.\ D {\bf 50}, 846 (1994)
  [gr-qc/9403028].
 
 
 \bibitem{M3}
  S.~Alexander, A.~Marciano and L.~Modesto,
  Phys.\ Rev.\ D {\bf 85}, 124030 (2012)
  [arXiv:1202.1824 [hep-th]].



\bibitem{M4}
  F.~Briscese, A.~Marciano, L.~Modesto and E.~N.~Saridakis,
  Phys.\ Rev.\ D {\bf 87}, 083507 (2013)
  [arXiv:1212.3611 [hep-th]].
  
  
  
\bibitem{kuzmin} 
  Y.~V.~Kuzmin,
  Sov.\ J.\ Nucl.\ Phys.\  {\bf 50}, 1011 (1989)
  [Yad.\ Fiz.\  {\bf 50}, 1630 (1989)].
  
  
\bibitem{HigherDG}
  A.~Accioly, A.~Azeredo and H.~Mukai,
  J.\ Math.\ Phys.\  {\bf 43}, 473 (2002);
  F.~d.~O.~Salles and I.~L.~Shapiro,
  arXiv:1401.4583 [hep-th].
  


\bibitem{Stelle} 
  K.~S.~Stelle,
  Phys.\ Rev.\ D {\bf 16}, 953 (1977).




  \bibitem{Shapirobook} I. L. Buchbinder, Sergei D. Odintsov, I. L. Shapiro,
  ``Effective action in quantum gravity", IOP Publishing Ltd 1992.
  
  
  
\bibitem{VN} P. Van Nieuwenhuizen,
Nuclear Physics B 60 478-492 (1973).

  
  
  
\bibitem{StableDeSitter1}
  T.~Biswas, A.~S.~Koshelev and A.~Mazumdar,
  Fundam.\ Theor.\ Phys.\  {\bf 183}, 97 (2016)
  [arXiv:1602.08475 [hep-th]].

  
\bibitem{StableDeSitter2} 
  T.~Biswas, A.~S.~Koshelev and A.~Mazumdar,
  Phys.\ Rev.\ D {\bf 95}, no. 4, 043533 (2017)
  [arXiv:1606.01250 [gr-qc]].

  
\bibitem{KKMR} 
  A.~S.~Koshelev, K.~Sravan Kumar, L.~Modesto and L.~Rachwal,
  arXiv:1710.07759 [hep-th].

  
  
\bibitem{shapiro3} 
  L.~Modesto and I.~L.~Shapiro,
  Phys.\ Lett.\ B {\bf 755}, 279 (2016)
  [arXiv:1512.07600 [hep-th]].  
  
\bibitem{HigherDG0} 
  F.~d.~O.~Salles and I.~L.~Shapiro,
  Phys.\ Rev.\ D {\bf 89}, no. 8, 084054 (2014)
  [Phys.\ Rev.\ D {\bf 90}, no. 12, 129903 (2014)]
  [arXiv:1401.4583 [hep-th]].
  
  
\bibitem{betaGN} L.~Modesto, L.~Rachwal and I.~L.~Shapiro, \textit{Renormalization group in super-renormalizable quantum gravity}, [arXiv:1704.03988 [hep-th]].
  
  \bibitem{Efimov} G. V. Efimov, ``Nonlocal Interactions" [in Russian], Nauka, Moscow (1977). 


\bibitem{GBV} 
A. O. Barvinsky and Vilkovisky, Phys. Rep. 119, 1 (1985) 1-74.


  \bibitem{Codello:2015oqa} 
  A.~Codello, R.~Percacci, L.~Rachwal and A.~Tonero,
 Eur.\ Phys.\ J.\ C {\bf 76}, no. 4, 226 (2016) [arXiv:1505.03119 [hep-th]]
  L.~Rachwal, A.~Codello and R.~Percacci,
 Springer Proc.\ Phys.\  {\bf 170}, 395 (2016).


   
    \bibitem{Nodimreg} 
  N.~Bao and T.~He,
  arXiv:1603.08531 [hep-th]. 




\bibitem{brisceseScalar} 
  F.~Briscese, E.~R.~B.~de Mello, A.~Y.~Petrov and V.~B.~Bezerra,
  Phys.\ Rev.\ D {\bf 92}, no. 10, 104026 (2015)
  [arXiv:1508.02001 [gr-qc]].

\bibitem{Nesterov:2010yi} 
  D.~Nesterov and S.~N.~Solodukhin,
  Nucl.\ Phys.\ B {\bf 842}, 141 (2011)
  [arXiv:1007.1246 [hep-th]].
  
  
  
\bibitem{ordinediretto} 
  D.~V.~Fursaev and G.~Miele,
  Nucl.\ Phys.\ B {\bf 484}, 697 (1997)
  [hep-th/9605153].
  



\bibitem{Fursaev:1994pq} 
D.~V.~Fursaev,
  ``Black hole thermodynamics and renormalization,''
 Mod.\ Phys.\ Lett.\ A {\bf 10}, 649 (1995)
  [hep-th/9408066].

\bibitem{Diaz:1989nx} 
  A.~Diaz, W.~Troost, P.~van Nieuwenhuizen and A.~Van Proeyen,
  Int.\ J.\ Mod.\ Phys.\ A {\bf 4}, 3959 (1989).

\bibitem{Myung} 
  Y.~S.~Myung,
  Phys.\ Rev.\ D {\bf 95}, no. 10, 106003 (2017)
  [arXiv:1702.00915 [gr-qc]].

%



\bibitem{conformal} L.~Modesto and L.~Rachwal, arXiv:1605.04173 [hep-th]. 



  \bibitem{shapiro2} 
  I.~L.~Shapiro,
  Phys.\ Lett.\ B {\bf 744}, 67 (2015)
  [arXiv:1502.00106 [hep-th]].  

 \bibitem{lm5} 
  L.~Modesto,
  Nucl.\ Phys.\ B {\bf 909}, 584 (2016)
  [arXiv:1602.02421 [hep-th]].  


%
%


  
  
  

  

 





%
%
   
   



  

%
%
%
  
  
    

  
  
    


  
  

  
    
  
  
\bibitem{revnl} L.~Modesto and L.~Rachwal, 
Int.\ J.\ Mod.\ Phys.\ D {\bf 26}, 1730020 (2017).


\bibitem{CalcagniStability}
  G.~Calcagni and L.~Modesto,
  Phys.\ Lett.\ B {\bf 773}, 596 (2017)
  [arXiv:1707.01119 [gr-qc]].




\bibitem{Myung2} 
  Y.~S.~Myung and Y.~J.~Park,
  arXiv:1711.06411 [gr-qc].



\bibitem{completeness} C.~Bambi, L.~Modesto and L.~Rachwal, 
JCAP {\bf 1705}, no. 05, 003 (2017)  [arXiv:1611.00865 [gr-qc]].

\bibitem{singproc} L.~Modesto and L.~Rachwal, 
J.\ Phys.\ Conf.\ Ser.\  {\bf 942}, no. 1, 012015 (2017) [arXiv:1801.03193 [hep-th]].
  
\bibitem{KSno} 
  A.~S.~Koshelev and A.~Mazumdar,
  Phys.\ Rev.\ D {\bf 96}, no. 8, 084069 (2017)
  [arXiv:1707.00273 [gr-qc]].

  
  
 
  



  

  











\end{thebibliography}
\end{document}